\documentclass[final,5p,times,twocolumn]{elsarticle}

\usepackage{graphicx}
\DeclareGraphicsExtensions{.eps,.ps,.eps.gz,.ps.gz} 

\usepackage{amssymb}
\journal{Nuclear Physics B}

\begin{document}

\begin{frontmatter}



\title{A Fast General-Purpose Clustering Algorithm Based on FPGAs for High-Throughput Data Processing}

\author[LNF]{A. Annovi\corref{cor}}
\ead{alberto.annovi@lnf.infn.it}
\author[LNF]{M. Beretta}
\ead{matteo.beretta@lnf.infn.it}
\cortext[cor]{Corresponding author}
\address[LNF]{INFN - Laboratori Nazionali di Frascati, via E. Fermi 40, Frascati}

\begin{abstract}
We present a fast general-purpose algorithm for high-throughput clustering of data "with a two dimensional organization". The algorithm is designed to be implemented with FPGAs or custom electronics.  The key feature is a processing time that scales linearly with the amount of data to be processed. This means that clustering can be performed in pipeline with the readout, without suffering from combinatorial delays due to looping multiple times through all the data. This feature makes this algorithm especially well suited for problems where the data has high density, e.g. in the case of tracking devices working under high-luminosity condition such as those of LHC or Super-LHC.

The algorithm is organized in two steps: the first step (core) clusters the data; the second step analyzes each cluster of data to extract the desired information. The current algorithm is developed as a clustering device for modern high-energy physics pixel detectors. However, the algorithm has much broader field of applications.
In fact, its core does not specifically rely on the kind of data or detector it is working for, while the second step can and should be tailored for a given application. For example, in case of spatial measurement with silicon pixel detectors, the second step performs center of charge calculation. Applications can thus be foreseen to other detectors and other scientific fields ranging from HEP calorimeters to medical imaging.

An additional advantage of this two steps approach is that the typical clustering related calculations (second step) are separated from the combinatorial complications of clustering.  This separation simplifies the design of the second step and it enables it to perform sophisticated calculations achieving offline-quality in online applications. The algorithm is general purpose in the sense that only minimal assumptions on the kind of clustering to be performed are made.

\end{abstract}


\end{frontmatter}


\section{Introduction}
%%\label{}

Pixel detectors have an extremely broad spectrum of applications,
ranging from high-energy physics to the photo cameras of everyday life. 

A large fraction of applications can benefit from a high-through clustering device that processes the collected data on the fly.
It can serve as a first data-processing step with several purposes.
Pixel clustering is useful to reduce the amount of data at an early stage. This is extremely important when the data rate to process is large.
The information from all pixels in the cluster can be summarized by the cluster properties of interest.
For the detection of ionizing particle, the output would be the best estimate of the particle spatial position.
Medical applications that search for clusters of anomalous density in medical images~\cite{MedicalClustering} can also benefit from a fast clustering device.

We propose a general purpose clustering algorithm that achieves high-throughput while maintaining the flexibility needed to be adaptable to different applications.

This algorithm is being developed as part of the Fast Tracker~\cite{FTK} trigger upgrade for the ATLAS experiment~\cite{ATLAS}.
It has many potential applications.
For high-energy physics tracking detectors, a fast clustering device allows us to imagine a self-clustering detector, where collected data is analyzed and clustered on the fly.
The data transmitted by the thousands of fibers could be received and analyzed by a clustering device located on the remote end of the fibers (off-detector).
An even more challenging possibility would be to have the clustering logic integrated on the detector front-end itself.
In this case we could take advantage of the data reduction associated with clustering to either reduce the power consumption need to send data off-detector or to increase the output rate.

\section{The Fast Tracker application}
%%\label{}

The first application is clustering of the ATLAS pixel detector data as part of the Fast Tracker processor.
The Fast Tracker is an upgrade for the ATLAS trigger system. It will analyze Silicon tracker data at the L1 output rate. The Fast Tracker will provide reconstructed tracks as input to the High Level Triggers~\cite{HLT} in alternative to raw detector hits.
Track reconstruction will cover the entire Inner Detector with offline-quality.
In order to achieve these goals the first step of the Fast Tracker processing needs to perform clustering of the pixel detector data.

The main challenge is to process the $\approx$160Gbits input data rate. Data is received over 132 S-Link fibers each running at a 1.2 Gbits.
The data format consist of 32 bits words at a rate of 40MHz where each word corresponds to one detector hit,  which encodes hit coordinates (row and column) within the module and the Time over Threshold (ToT) information.
Through this document we will use the ability to process hits at a 40MHz rate as a benchmark.
The clustering algorithm must provide high-quality resolution, in order to match the high quality needed the High Level Triggers.  

\subsection{Pixel module readout order}
The ATLAS pixel modules are described in ref.~\cite{PIXEL}. Each module is readout by 16 Front End chips (FE) organized in two rows of 8 FE each. Each FE reads out pixel hits by double column. Because of its logic, hits in a double column are scrambled in the readout, while packets of data from each double column are read out one after the other in a fixed order. Data packets from each of the 16 FE chips is received by the MCC chip that performs event building. The data stream after MCC processing contains data from each FE chips organized in consecutive packets.
In other words, the final data stream sent out by the module has packets of data corresponding to double columns which are readout in a sorted and predefined order one after the other. The data is scrambled only within each double column.

\subsection{The 2D problem}

In the case of silicon strip detectors, the clustering becomes very easy if the strip readout is ordered. In fact, in this case contiguous hits belonging to the same cluster are also contiguous in the readout data stream.

For pixel detectors, because of the intrinsic 2D nature of the problem, it is impossible to define a readout algorithm such that all hit pixels belonging to one cluster are readout consecutively. This means that the algorithm somehow has to loop over the list of hit pixels in a module in order to recognize that two or more hits belong to the same cluster. 
We define a cluster as a set of hits that are contiguous either through a common side or a common vertex (in order words diagonally).
If we imagine a software algorithm looping over all data from a pixel module, we can estimate that the algorithm complexity would be
proportional to the square of the number of hits.

The algorithm would take advantage of readout ordering to achieve a complexity that is proportional to the square of the local occupancy.
In any case, the computing time will scale more than linearly with occupancy.
This leads to potential troubles at high instantaneous-luminosity when the hit occupancy due to many pile-up events is high.

Is it possible to design a clustering algorithm that, regardless of the detector occupancy, has a processing time that scales linearly with the number of hits? This is a very desirable feature because it matches the readout time that is always linear with the number of hits.
In practice any algorithm that runs with a processing time that is linear with the number of hits, would be able to keep the pace with the readout at any instantaneous-luminosity.
We will show that a simple hardware base algorithm can satisfy this requirement.

\subsection{The algorithm}
\begin{figure}[!ht]
  \begin{center}
    \resizebox{0.48\textwidth}{!}{\includegraphics{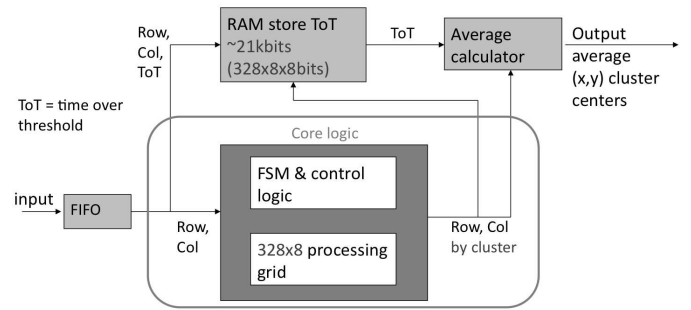}}
    \caption{\label{fig:block_diagram}Block diagram of the clustering algorithm.}
  \end{center}
\end{figure}

The goal is an algorithm complexity that scales linearly with the number of hits, while achieving offline resolution.
In order to combine speed with quality, we split the algorithm in two steps.
The first step will group together hits belonging to the same cluster.
This is the critical part of the algorithm where we will exploit a dedicated logic in order to achieve linear processing time.
The second step will analyze hits in the cluster in order to calculate the cluster properties of interest.
This second step, being separate from the first one, will have the flexibility needed to achieve the best resolution.
As an additional benefit, it will be easy to specialize it to suit different applications.
Fig.~\ref{fig:block_diagram} shows the two algorithm steps working together.
The first step is represented by the core logic. The second step is represented by the average calculator.
A detailed description of the first step follows.

\begin{figure}[!ht]
  \begin{center}
    \resizebox{0.48\textwidth}{!}{\includegraphics{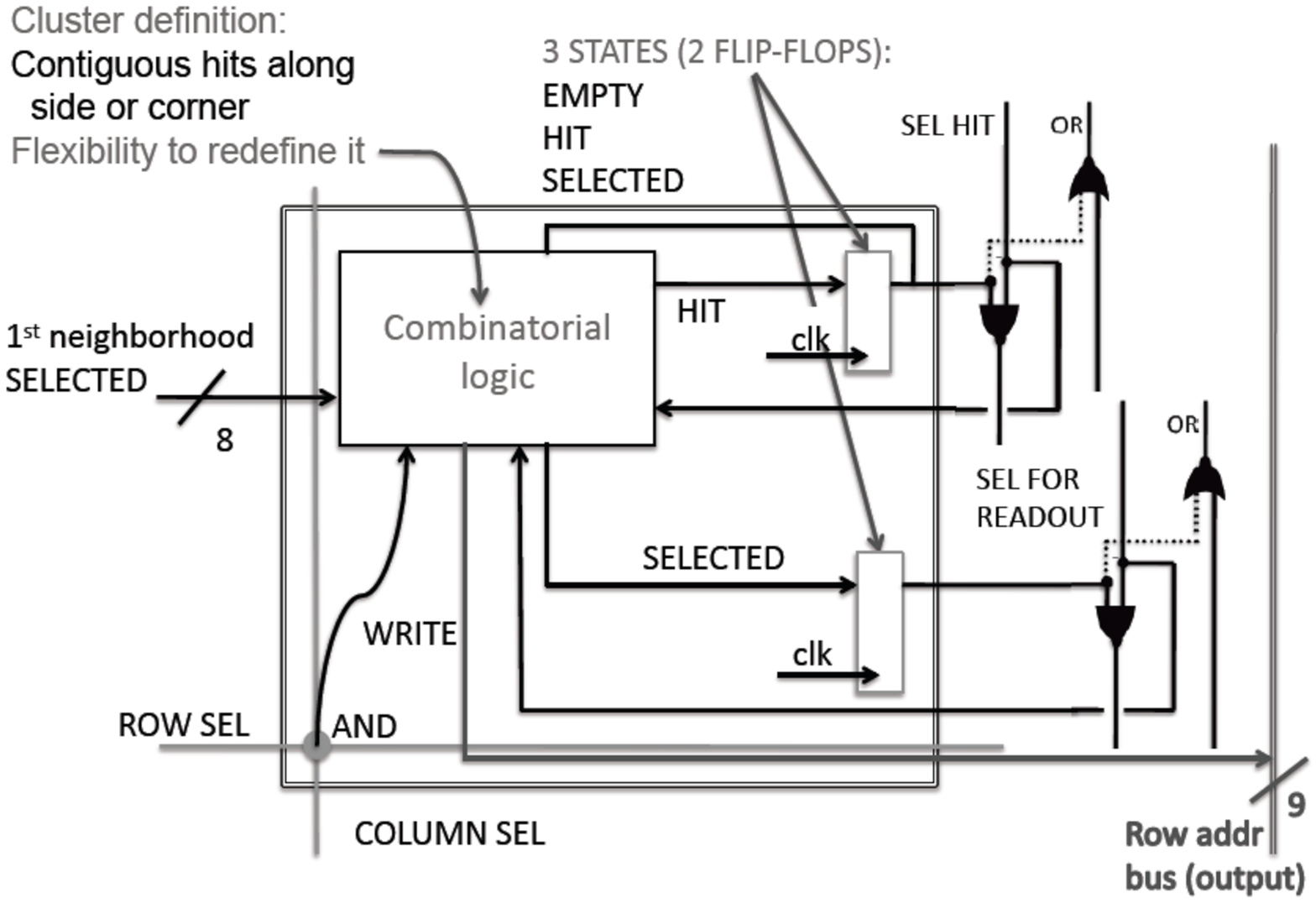}}
%%    \resizebox{0.48\textwidth}{!}{\rotatebox{90}{\includegraphics{CellLogic.eps}}}
    \caption{\label{fig:cell}Logic of the elementary clustering cell.}
  \end{center}
\end{figure}

In order to achieve linear processing time, we exploit the powerful logic of FPGA.
In fact, the 2D structure of FPGAs is suitable for mapping the 2D structure of a pixel module.
So let's imagine that the FPGA logic represents a 2D grid of clustering cells. Fig.~\ref{fig:cell} shows the logic of the elementary cell.
Each cell has three possible states: EMPTY, HIT, SELECTED.
The first step of the algorithm will receive data from one pixel module and start to load it onto the FPGA grid, by marking the corresponding cells as HIT (WRITE signal in the figure). Once all data is loaded in the grid, an external Finite State Machine (FSM) will select the first hit in the grid through a priority chain. The first hit will be marked as SELECTED. The SELECTED status propagates to nearby HIT cells through local logic. This is the key step of the algorithm that avoids any loop. As soon as the first cell is SELECTED the FSM will start to readout all selected cell positions. Again the FSM will start reading out SELECTED cells using another identical priority chain. The list of these positions will be output and it will represent the first cluster. Please note that the propagation of the SELECTED status and the readout can happen at the same time with the loss of just one clock cycle needed to SELECT the seed cell. At this point the algorithm can start over selecting next first hit in the grid, and so on.
The two priority chains match the readout order of data that is sorted in the direction perpendicular to the columns.
The column index will be the most significant and the row index the least significant. Hence, the first hit is the one in the column with the lowest column index and within the same column the hit with the lowest row index.
In fig.~\ref{fig:cell} the definition of a cluster is encoded in the combinatorial logic box.
Here we have the flexibility to redefine when hits belong to the same cluster without changing the overall structure of the algorithm.

The description above defines the core (first step) of the algorithm. A second algorithm step will calculate the center of the cluster. The center will be calculated as a ToT weighted average of the positions of the hits. 
This can be done with high speed and limited hardware resources after the cluster has been identified by the first algorithm step.
The second step is completely decoupled from the first one. It can be modified to calculate any cluster property of interest depending on the specific application.
In the description above the FPGA grid does not store the ToT information. The ToT is stored in a RAM while hits are being loaded on the grid. During grid readout the ToT will be easily retrieved and sent to the second step logic.

The algorithm described above uses a 2D grid of logic cells that represents a whole pixel module. It is 328x144 cells wide.
It is important to evaluate the hardware size and the clock speed of this logic.
We anticipate that the whole grid will need a lot of resources. Then we will describe how to reduce the amount of logic to a manageable size.

For this exercise, we implemented the logic on a xv5vlx330 Xilinx FPGA~\cite{Xilinx}. The xv5vlx330 is currently the largest of the xv5vlx group. We use it as a reference point. 

\begin{table} 
  \begin{center} 
    \begin{tabular}{ccc}
      grid size &	clock period & area usage \\ \hline
      8x8 &	 6ns	& \~1\% \\
     120x8 &	 13ns&	 5\% \\
     32x32	 & 13ns&	 6\% \\
     64x32	 & 16ns&	 11\% \\
     256x8	 & 15ns&	 11\% \\
     328x8	 & 17ns&	 16\% \\
     120x32 &	 20ns&	 21\% \\
    \end{tabular} 
    \caption{\label{tab:algo_performances}Algorithm performances on a xc5vlx330 FPGA.}
  \end{center}
\end{table}

Tab.~\ref{tab:algo_performances} reports the clock period and area size as function of the grid size. The area usage scales as expected with the area of the grid. Scaling the 328x8 to the full pixel module we obtain an area of 280\% of one xv5vlx330 FPGA.
This clearly indicates that we need a more efficient way to cluster hits. A solution to this problem is in the next section.
The clock speed also scales with the gird size. The priority logic is the net with the maximum delay. It determines the minimum clock period. Because the priority logic involves all cells in the grid, the clock period scales with the grid size. 

\subsection{Using a sliding grid to reduce hardware usage}
We have seen that using an FPGA-based processing grid that reproduces a full pixel module is not feasible. It would also be very inefficient. In fact, the current algorithm loads all data on the grid.,then it starts clustering data from the first (priority wise) hit. It is clear that pixels that are far away from the first hit do not take part in the clustering process. It is thus useless to have a grid as big as the module.

What is the minimum size of the processing grid that we can use? Of course it must be larger than a cluster. Most clusters are up to 3 pixel wide along the $\phi$ direction and up to 5 pixel wide along the z\footnote{$\phi$ and z coordinates refer to the cylindrical geometry of pixel modules in the barrel.} direction~\cite{PixelTDR}.
Pixels are readout in groups of double columns, i.e. scrambled within a double column but sorted one double column after the other. Because of this, the grid must be at least as long as a double column along the $\phi$ direction. This is needed in order to have room to store all hits received during readout for a given double column. An alternative option would be to store separately (e.g. in a memory) hits that are away by more than 3 pixel w.r.t to the first (priority wise) hit. However this would require checking all hits multiple times, with a price on processing time. This could lead to further optimization of the algorithm. 
One column is 164 pixels long.  It corresponds to one FE chip. So our grid must be 164 pixels long along $\phi$. We chose to make it 328 pixels wide in order to avoid edge effects at the center of the module. Along the z direction we must choose a value bigger than 5. A good number is 8 that allows for extra some margin.

\begin{figure}[!ht]
  \begin{center}
%%    \resizebox{0.48\textwidth}{!}{\rotatebox{90}{\includegraphics{SlidingLogic_BW.eps} } } %% 3.png
    \resizebox{0.48\textwidth}{!}{\includegraphics{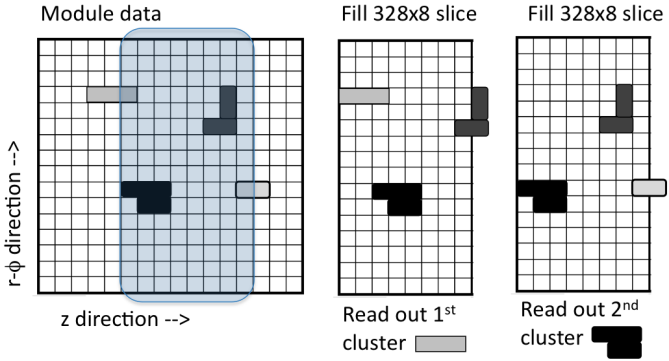}}
    \caption{\label{fig:sliding} Example of sliding window logic at work.}
  \end{center}
\end{figure}

We can use a sliding window algorithm that scans the pixel module from left to right and clusters all hits. I assume a window size of 328x8 pixels. 
Fig.~\ref{fig:sliding} represents the working principle of the sliding window. The left most diagram shows a pixel module with all its hits and clusters. Those are received from the input S-Link and stored on the FIFO.
The center diagram shows the sliding window a 328x8 grid of processing logic. When we start to process a new module the sliding grid will be empty. As soon as the first hit is received by the FIFO, the FSM will know its column number. Thus we can align the left edge of the sliding grid to this column (center diagram). The alignment procedure is virtual it corresponds to labeling the first column of the sliding grid. In other words the column number of the first hit is stored in a register. Please note that if the column number is the second of a double column, we should align the sliding grid one column to the left of the first hit in order to allow room for all hits in the double column.
The algorithm starts loading hits in the grid until the first hit beyond the sliding grid width is received.
Of course a fraction of the hits and even part of some clusters will not be loaded at this point.
At this point load phase ends and one cluster is readout.
 The first (priority wise) hit is SELECTED for readout. For the next cycle we can start reading out all selected hits. The SELECTED signal is propagated from the first hit to its own neighborhood pixels that are hit.
After one cluster is fully readout, the algorithm starts over aligning the grid to the next first hit and the grid is ready to load more hits.
This is shown in the right diagram. Hits that were left out during the previous load phase can now be loaded on the grid.

This is the list of algorithm steps:
\begin{itemize}
\item align sliding grid to first hit
\item load hits
\item SELECT first (priority wise) hit
\item readout all SELECTED hits
\item start over with next cluster
\end{itemize}

This algorithm correctly clusters hits with the only exception of clusters exceeding 8 columns in length that will be split. This algorithm has a manageable hardware size. It can be implemented with 15\% area usage on a xc5vlx330 or with 30\% area usage on the smaller xc5vlx155. From the list of steps above we can extract that the algorithm will need 2 clock cycles to process each hit (one for loading it and one for reading it out) plus 2 clock cycles per cluster (one for aligning the grid and one for SELECTING the first hit). The clock cycle counting has been verified with FPGA simulation. If we assume clusters of 2 hits on average, we get 3 clock cycles per hit. For larger clusters the average number of clock cycles per hit is smaller. At this point we can compare the hit processing time of 15ns (clock period) times 3 clock cycles per hit that equals 45ns with the hit rate from S-Link input that is 40MHz or 25ns. This means that we need to gain a factor of 2 in speed. It can be gained using two sliding windows processing two modules in parallel. This would result in an area usage of 60\% on a xc5vlx155 FPGA.

\subsection{The second algorithm step}
The implementation of the second algorithm step becomes easy because it is decoupled from the first step. We tried two options. 
A first option that ignores the ToT information. In this simplified case the resolution is still good.
The residual resolution with respect to the offline cluster position has an RMS  of a tenth of a pixel along each direction. 
As second option we used the full ToT information in the same way as the current offline algorithm does.
In this case we achieve the same resolution as the offline with the exception of a few anomalous clusters that exceed 8 columns in width.

\section{Conclusions}
We have developed and studied a clustering algorithm for the pixel detectors. The proposed algorithm achieves linear processing time with respect to the number of readout hits. For this reason it is intrinsically stable with respect to detector occupancy. The algorithm can be implemented in hardware with one xc5vlx155 FPGA. Is uses 60\% of its logic in order to process data from one input S-Link.
The algorithm is flexible in the definition of clustering and in the calculation of output cluster properties. It can be adapted to suit most applications.

\section*{Acknowledgments}
We thank Mauro Dell'Orso, Paolo Laurelli, Giovanni Maccarrone and Andrea Sansoni for the fruitful discussion on this subject.

\end{document}